\documentstyle[12pt]{article}  
\advance\hoffset by -7mm  
\makeatletter
\@addtoreset{equation}{section}
\makeatother

\renewcommand{\title}[1]{\null\vspace{25mm}

\noindent{\Large{\bf #1}}\vspace{10mm}

\noindent {\large By }}
\newcommand{\authors}[1]{\noindent{\large #1}\vspace{3mm}

}
\newcommand{\address}[1]{\noindent #1\vspace{5mm}

}
\renewcommand{\abstract}[1]{\vspace{19mm}

\noindent{\small{\em Abstract.} #1}\vspace{2mm}

} 

\begin{document}
\title{Creation of multiple de Sitter universes\\ inside 
                a Schwarzschild black hole}
\authors{C.Barrab\`es }
\address{EP 93 C.N.R.S., Dept.Physique, Facult\'e des Sciences,   
        37200 Tours, France}
\authors{and V.P.Frolov }
\address{C.I.A.R. Cosmology Program, University of Alberta,     
        Edmonton, Canada, T6G 2J1}
\abstract{A classical model for the interior structure of a Schwarzschild  
        black hole which consists in creating multiple de Sitter
        universes with lightlike boundaries is proposed.The interaction
        of the boundaries is studied and a scenario leading to
        disconnected de Sitter universes is described.}

   The interior structure of a black hole and its final state after 
   evaporation are still two intriguing problems.Their study requires
   the knowledge of physics at Planckian scales and it is generally 
   believed that only a junction of quantum theory to gravity will
   provide a proper solution.Quantum effects become important as 
   one penetrates deeply inside the horizon of a black hole and
   it is hoped that they will prevent the formation of a singularity
   which,in classical general relativity, is the unavoidable end
   state of gravitational collapse according to Penrose's theorem
   (let us recall that a singular behavior already happens
   at the Cauchy horizon of a charged or rotating black hole). 

   Following these ideas a classical singularity-free model for the internal
   evolution of a Schwarzchild black hole was proposed by Frolov,
   Markov and Mukhanov (FMM) \cite{FMM} a few years ago.It relies upon
   two assumptions-i)limiting curvature hypothesis \cite{Mar} 
   -ii)transition to a
   de Sitter state in the final stage of gravitational collapse.
   Justifications of these two assumptions can be found in the 
   existence of corrective terms in the effective action for gravity,
   i.e. Polchinsky \cite{Pol},Mukhanov and Brandenberger \cite{MuBr}.
   It has also been shown \cite{PoIs} that the vacuum polarization 
   inside the horizon of a
   Schwarzschild black hole can have a self-regulation effect on the
   rise of curvature and that once the quantum fluctuations have died
   away the de Sitter state is the simplest possibility.For a Schwarzchild
   black hole with mass $m$ the curvature reaches Planckian values,
   $l_{Pl}^{-2}$, when the radial coordinate $r$ is of the order of
   $r_{0} = l_{Pl}(2m/l_{Pl})^{1/3}$ which corresponds to a value
   far inside the horizon but still in the classical regime,
   ($l_{Pl}<<r_{0}<<2m$) for a typical black hole.

   A key ingredient of the FMM model is that the transition from the
   vacuum phase to the false-vacuum phase occurs instantaneously
   along the homogeneous spacelike shell straddling the hypersurface
   $r=r_{0}$.One should however expect that this transition, which is
   induced by quantum fluctuations,occurs in a random way both in space
   and in time.A way of generalizing the FMM model which allows the
   existence of inhomogeneities consists in introducing the spontaneous
   creation of pairs of lightlike shells.A recent study \cite{BaIs} of 
   shells in the lightlike limit has shown that such a process is possible
   provided that some geometric matching relations are fulfilled.
   For static spherically symmetric spacetimes with metric of the form
   \begin{equation}
   ds^{2} = -f(r) dt^{2} + f(r)^{-1} dr^{2} + r^{2} (d\theta {^2} + sin\theta {^2} d\varphi {^2}
   \end{equation} 
   as it is the case for Schwarzschild and de Sitter,a pair of lightlike
   shells is created at the 2-sphere $r_0$ provided that  
   the functions $f$ to the past (here Schwarzschild,
   $f_{S}(r) = 1 - 2m/r$) and to the future 
   (here de Sitter,$f_{dS}(r) = 1 - r^2/a^2$) of the creation
   take identical values at $r_0$ ,see fig.1.
        \begin{figure}
        \let\picnaturalsize=N
        \def\picsize{13.0cm}
        \def\picfilename{jrfig1.eps}
        \ifx\nopictures Y\else{\ifx\epsfloaded Y\else\input epsf \fi
        \let\epsfloaded=Y
        \centerline{\ifx\picnaturalsize N\epsfxsize \picsize\fi
        \epsfbox{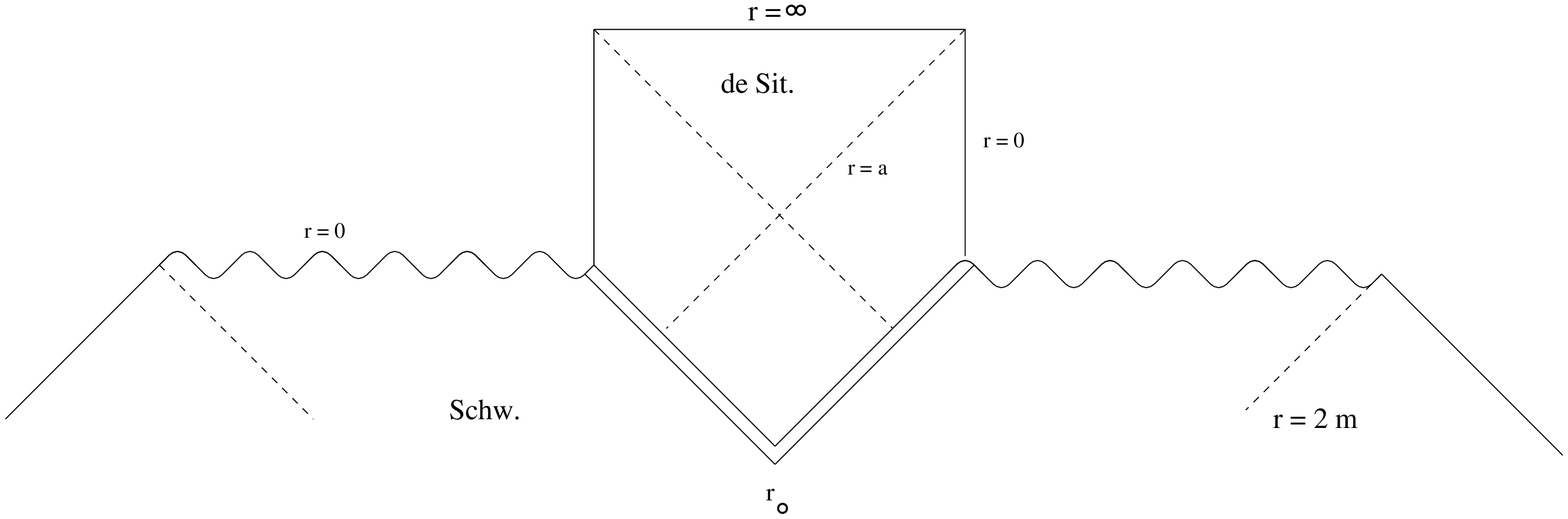}}}\fi
        \caption{Creation of a pair of lightlike shells at  
        a 2-sphere $r_{0}$.These shells form the boundaries 
        of a de Sitter closed universe in the future of the creation.}
        \end{figure}

   This implies the following matching relation
   \begin{equation}
   r_{0} = a (2m/a)^{1/3}
   \end{equation}
   where the de Sitter parameter $a$ is of the order of the Planck length.
   Immediately after their creation the two lightlike shells form the
   boundaries of a closed de Sitter universe.Their surface energy
   density is zero at $r_{0}$ and becomes increasingly negative 
   as energy is pumped to create the false-vacuum phase.

   The process of creation of one de Sitter bubble which we have
   just described can be repeated at any point of spacetime
   where the curvature becomes planckian.If now two de Sitter 
   bubbles are created closely enough their
   lightlike boundaries may intersect and various scenarios can be
   imagined to occur at the intersection.An interesting 
   one leading to the formation of two disconnected de Sitter
   universes corresponds to the case when the two ingoing lightlike
   shells merge into a single timelike shell separating the two 
   de Sitter universes (fig.2).
        \begin{figure}
        \let\picnaturalsize=N
        \def\picsize{11.0cm}
        \def\picfilename{jrfig2.eps}
        \ifx\nopictures Y\else{\ifx\epsfloaded Y\else\input epsf \fi
        \let\epsfloaded=Y
        \centerline{\ifx\picnaturalsize N\epsfxsize \picsize\fi
        \epsfbox{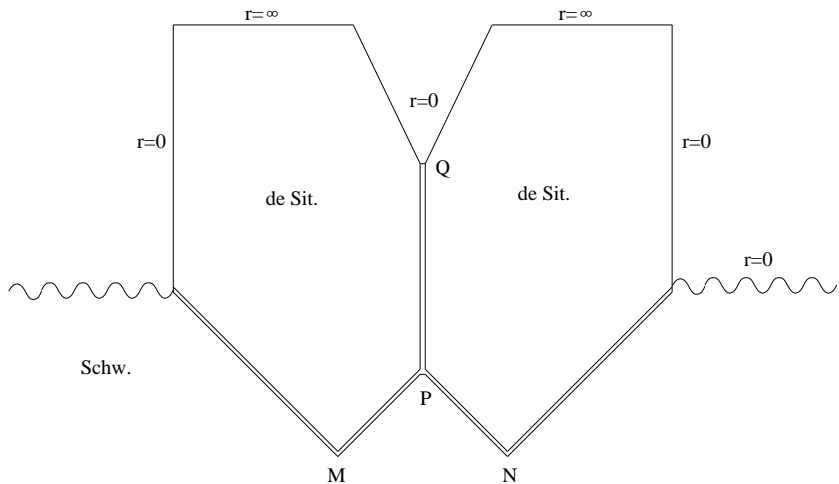}}}\fi
        \caption{Merging into a single timelike shell
        at the 2-sphere P,$r_{1}$,
        of the two lightlike shells forming the boundaries
        of the two de Sitter bubbles which were initially 
        created at M and N.The timelike shell contracts to
        zero radius at Q and the de Sitter universes get
        disconnected.}
        \end{figure}

   Such a scenario is realized
   provided that matching relations between the geometry of
   the adjacent spacetimes are satisfied and also initial conditions
   for the timelike shell are given.If the intersection takes place
   at the 2-sphere $r_{1}$ with $r_{1}<r_{0}$, then the initial velocity
   $\dot{r}_{1}$ and the initial inertial mass $M(r_{1})$ of the timelike
   shell have to be equal to
   \begin{equation}
   \dot{r}_{1} = -[4 f_{S}(r_{1})]^{-1} [ f_{dS}(r_{1}) + f_{S}(r_{1}) ]^{2}
   \end{equation}
   \begin{equation}
   M(r_{1}) = 2 m(r_{1}) \mid f_{dS}(r_{1})\mid ^{-1/2}
   \end{equation}
   where $m(r_{1})$ is the surface-energy density of the lighlike shells
   at the intersection.

   The future evolution of the timelike shell is governed by the
   following equation
   \begin{equation}
   \dot{r}^2 + V(r) = -1
   \end{equation}
   where the effective potential $V(r)$ is equal to
   \begin{equation}
   V(r) = -r^2/a^2 - 4\pi r^2 \sigma^2
   \end{equation}
   $\sigma$ being the surface-energy density of the timelike shell.
   As a result of the above equations the colliding de Sitter
   bubbles can either remain connected or get disconnected.The
   condition to get their separation is that 
   they are created at a spatial distance $\Delta t$ being 
   larger than some minimum value $\Delta t_{min}$ equal to 
   \begin{equation}
   \Delta t_{min} = a (a/m).
   \end{equation}
   As $a \sim l_{Pl} \ll m$,the minimum value $\Delta t_{min}$ 
   is very small
   and one may conclude that this scenario preferably produces 
   disconnected universes.

   It is interesting to compare the maximum number $N_{max}$
   of disconnected closed de Sitter universes which can be
   created during the time of evaporation $T_{evap}$ of the
   black hole to the number $N^{H}$ of emitted quanta of the
   Hawking radiation.As $T_{evap} \sim t_{Pl} (m/m_{Pl})^{1/3}$,
   where $t_{Pl}$ and $m_{Pl}$ are the Planck time and mass resp.,
   at most $N_{max} \sim (m/m_{Pl})^4$ new de Sitter universes
   can be formed.For typical black holes this number is much larger than
   $N^{H} \sim (m/m_{Pl})^2$  and the information which is lost
   during the evaporation of the black hole could have enough available
   space to escape.This remark might be of interest to the
   information-loss puzzle.
   
   The creation of de Sitter universes which has been described in this
   work only concerns a Schwarzschild black hole.It could also be
   applied to a charged or/and rotating black hole.However 
   it has been shown in these two cases that,due to the 
   mass inflation phenomenon \cite{Inf} the curvature of
   spacetime becomes infinite near the Cauchy horizon.
   Therefore our model should be applied at this place instead
   of the vicinity of the singularity $r = 0$.

\end{document}